\documentclass[12pt, draftclsnofoot, onecolumn]{IEEEtran}
\usepackage{enumerate}
\usepackage{graphicx,indentfirst}
\usepackage{booktabs}
\usepackage{indentfirst}
\usepackage{amsmath}
\usepackage{cite}
\usepackage{stfloats}
\usepackage{multicol}
\usepackage{makecell}
\usepackage{multirow}
\usepackage{subfig}
\usepackage[english]{babel}
\usepackage{amsthm}
\usepackage{bm}
\usepackage{amsmath}
\usepackage{multirow}
\usepackage{paralist}
\usepackage{algorithm, algorithmic}
\usepackage{amssymb}
\usepackage{amssymb}
\usepackage{mathrsfs}
\usepackage{amsfonts}
\usepackage{color}

\theoremstyle{plain}

\IEEEoverridecommandlockouts
\begin{document}
\bibliographystyle{IEEE2}

\title{When Mobile Blockchain Meets Edge Computing}
\author{Zehui~Xiong,~\IEEEmembership{Student Member,~IEEE,} Yang~Zhang,~\IEEEmembership{Member,~IEEE,} Dusit~Niyato,~\IEEEmembership{Fellow,~IEEE,} Ping~Wang,~\IEEEmembership{Senior Member,~IEEE,} and~Zhu~Han,~\IEEEmembership{Fellow,~IEEE}\\
\thanks{Zehui Xiong, Dusit Niyato and Ping Wang are with School of Computer Science and Engineering, Nanyang Technological University, Singapore. Yang Zhang is with School of Computer Science and Technology, Wuhan University of Technology, China. Zhu Han is with the University of Houston, Houston, TX 77004 USA, and also with the Department of Computer Science and Engineering, Kyung Hee University, Seoul, South Korea.}\vspace*{-4mm}}
\maketitle

\begin{abstract}
Blockchain, as the backbone technology of the current popular Bitcoin digital currency, has become a promising decentralized data management framework. Although blockchain has been widely adopted in many applications, e.g., finance, healthcare, and logistics, its application in mobile services is still limited. This is due to the fact that blockchain users need to solve preset proof-of-work puzzles to add new data, i.e., a block, to the blockchain. Solving the proof-of-work, however, consumes substantial resources in terms of CPU time and energy, which is not suitable for resource-limited mobile devices. To facilitate blockchain applications in future mobile Internet of Things systems, multiple access mobile edge computing appears to be an auspicious solution to solve the proof-of-work puzzles for mobile users. We first introduce a novel concept of edge computing for mobile blockchain. Then, we introduce an economic approach for edge computing resource management. Moreover, a prototype of mobile edge computing enabled blockchain systems is presented with experimental results to justify the proposed concept.
\end{abstract}

\newpage

\section{Introduction}
Blockchain works as a decentralized public ledger which can store data, i.e., records of transactions. Blockchain outperforms centralized ledger approaches which suffer from low efficiency due to a bottleneck, single point failure and security attack, and moral hazard. Alternatively, data is recorded by blockchain as blocks, e.g., collections of transactions, which form a linked list data structure to indicate logical relations among the data added to the blockchain. No centralized entity or intermediary is required to maintain data blocks. Instead, the data blocks are copied and shared over an entire blockchain network to prevent system failure, data manipulation, and cyber attacks. Blockchain has been applied in several distributed system scenarios, e.g., content delivery networks~\cite{Herbaut.Video.2017} and smart grid systems~\cite{Kang.PHEVBlockchain.2017}.

The key concept to guarantee data integrity and validity in blockchain is a computational process defined as mining~\cite{Pass.FruitChain.2017}. To append a new block of data to the current blockchain, a blockchain user, i.e., miner, is required to solve a compute-intensive proof-of-work (PoW) to obtain a hash value that links previous block to the current block. After solving the PoW, the result is broadcast to other miners in the networks for validation. The new block is successfully added if majority of miners agree or reach consensus. Many consensus protocol gives a reward to the successful miner as an incentive to solve PoW. However, due to the required PoW computation contribution, resource-limited nodes such as Internet of Things (IoT) and mobile devices cannot directly participate in the mining and consensus process, introducing major challenge in blockchain applications for IoT and other mobile services. 

Mobile edge computing~\cite{Abbas.MECSurvey.2017} (MEC) architecture has been introduced to leverage available computing power in mobile environments. Local data centers and servers are deployed by a service provider at the ``edge'' of mobile networks, e.g., base stations of radio access networks. MEC is the key technology to meet stringent low-latency requirements of 5G networks~\cite{Vicent.2017.5G}. Mobile devices can access the edge servers to enhance their computing capability, e.g., IoT sensing data processing. With this feature, edge computing becomes a promising solution for mobile blockchain applications, the benefits of which are as follows. Firstly, by incorporating more miners, the robustness of the blockchain network is naturally improved. Secondly, the mobile users have an incentive from the reward obtained in the consensus process. However, edge computing services are deployed by the provider to maximize its benefit. As such, a pricing issue of the edge services arises. Accordingly, given the pricing adopted by the edge computing service provider, miners also need to optimize their demand for edge computing service for solving PoW to maximize their payoffs. Pricing is a commonly used incentive technique adopted in wireless networks~\cite{Zhang.Fog.2017}. For example, in~\cite{YingG.Pricing.2016}, the authors proposed a pricing mechanism to incentivize cooperative communication between mobile terminals, which can significantly decrease both the communication and battery outage.

In this article, we consider an edge computing enabled mobile blockchain network, where IoT devices or mobile users can access and utilize resources or computing services from an edge computing service provider~\cite{Zhang.Contract.2016} to support their blockchain applications. First, we present overviews of blockchain and edge computing architecture in Section~\ref{Sec:PrelimBlockchain} and Section~\ref{Sec:MECBlockchain}, respectively. We propose a prototype of an edge computing system for mobile blockchain in Section~\ref{Sec:Demo}. Then, in Sections~\ref{Sec:Model} and \ref{Sec:Numerical}, we propose pricing schemes for the edge computing services for mobile blockchain. Some typical numerical results are presented to show important findings from the proposed pricing scheme.

\section{Blockchain Overview}	\label{Sec:PrelimBlockchain}
The basic idea of blockchain is a chain-shaped data structure, also called a chain of blocks or a blockchain. Each data block contains historical verified data or transactions. The blockchain can be replicated and spread to all participants, i.e., users, in a publicly accessible blockchain network such that it is synchronized globally.

	\subsection{Blockchain Architecture}	\label{Subsec:Arch}
	{\em Data Organization in Blockchain:}
	Figure~\ref{Fig:edge} shows the diagram of a blockchain structure. Each block in the blockchain typically contains two parts, i.e., transaction data and hash values. Transaction data are recorded by the blockchain users or systems, e.g., mobile sensing devices. Hash values are used to store coded or secured information. A hash value in a block is generated based on the information of the previous block, which is similar to a link pointing from the current block to the block prior to it. The first block in the blockchain is called genesis block. A newly generated block is validated and sequentially appended to the genesis block or the block earlier to form a chain of blocks. 

	{\em Basic blockchain implementation:} In general, blockchain operates in the following steps.
		\begin{enumerate}
			\item A blockchain user performs a transaction and creates a new transaction record. The new transaction will be transferred to neighboring peer users in the blockchain network.
			\item Each peer user collects the transferred transactions for a certain period of time. Invalid user transactions, e.g., fake transactions, will be discarded. After the period, the user which has collected a set of transactions packs the transactions into a block and performs mining by solving PoW. The users are referred to as miners.
			\item The mined block is then sent into the network to notify other users. Other users which receive the mined block will examine the validity with security mechanisms. If the block is proved valid by majority of users, it will be appended to the end of the current blockchain. This is the consensus process.
		\end{enumerate}

	\subsection{Consensus and Mining: A Trust Overlay on a Trustless Network}	\label{SubSec:Cons-and-Minn}
	{\em Distributed consensus:} Blockchain can offer more than cryptography for transactions and data in trustless distributed networks. While cryptographic techniques are adopted in mining to prevent the blockchain from being changed, falsified, or deleted by malicious middleman users, the blockchain is guaranteed to be acknowledged by many users. Consensus is a mechanism to ensure trust in the network, which means that users in the network commonly reach an agreement of a block added to the existing blockchain. The users can discover the false transaction injected by an attacker by obtaining and checking the acknowledged blockchain from the network. As such, the majority of users can ``veto'' to discard the false transaction. The deployment of consensus is useful for distributed computing systems with independent users, e.g., mobile networks~\cite{Eve.Schooler.2017}.
		
	{\em Proof-of-work based mining:} However, consensus may still be vulnerable to attacks, e.g., Sybil attack, where an attacker creates fake blockchain information and many pseudonymous users in the network. The fake blockchain information and pseudonymous users can lead to the consensus of false transactions generated by the attacker. The solution to such attacks is to raise the complexity of mining so that attackers cannot have enough computing power to support fake users in the network. Proof-of-work (PoW) is used to increase the mining complexity. It is a process of work which is difficult to produce but easy for verification. Solving PoW has to calculate a value that makes the header hash value lower than a given ``difficulty target''. Confirming and securing the integrity and validity of transactions are executed by a set of miners. In many blockchain systems, e.g., Bitcoin, a miner which successfully mines a block receives the mining reward when the mined block is successfully added to the blockchain. The consensus guarantees the security and dependability of blockchain systems~\cite{Pass.FruitChain.2017}. In the blockchain network where mining is costly, attackers need to control over $51\%$ of all the computing power in the network to manipulate the blockchain. This is an enormous cost for the attackers which can rarely be feasible in practice.

\section{Edge Computing Services for Mobile Blockchain}	\label{Sec:MECBlockchain}
		\begin{figure}[htbp!]
		\begin{center}
		\includegraphics[width=.95\textwidth]{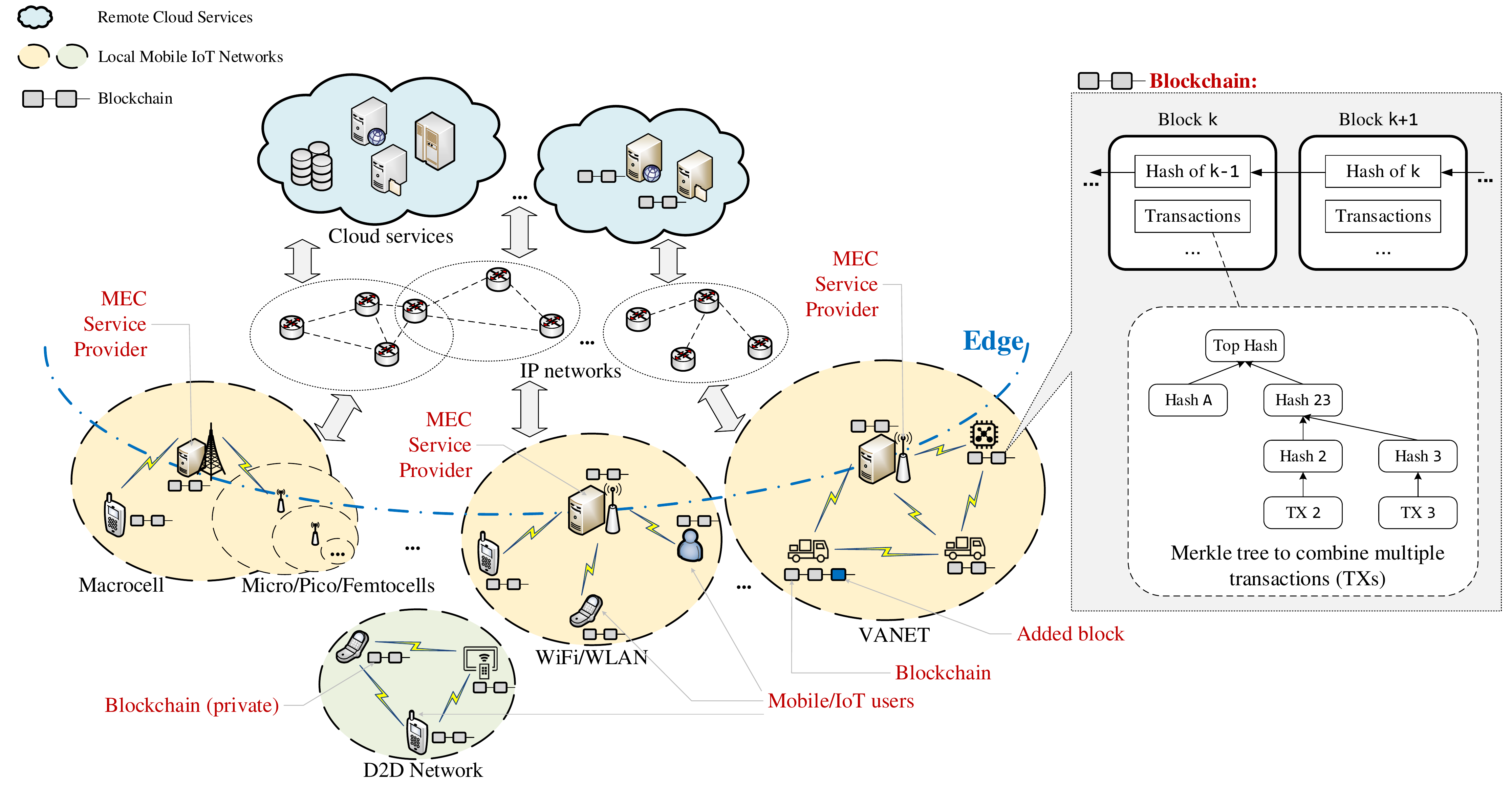}
		\caption{\small Mobile Edge Computing (MEC) Enabled Blockchain.}
		\label{Fig:edge}
		\end{center}
		\end{figure}
	
	\subsection{Beyond Bitcoin: Edge Computing Services for Mobile Blockchain}
	Recently, blockchain has found many applications in various networks and distributed systems today, one of them is IoT~\cite{Christidis.B4IoT.2016}. IoT systems connect a variety of physical objects such as mobile devices, sensors and actuators to the Internet. The IoT devices can sense, communicate, and exchange information with each other to achieve a certain goal of the systems, e.g., smart transportation, logistics, healthcare, and manufacturing. However, IoT devices are usually low-powered, geographically distributed, and possibly mobile. Limited computing resource and energy supply of IoT devices become major barriers when blockchain is applied to IoT systems specifically because of the mining process. Instead, mobile edge computing can supply computing and communication resources for mobile blockchain. Mobile edge computing~\cite{Abbas.MECSurvey.2017} allows service providers to deploy cloud computing services at the ``edge'' of the mobile Internet as shown in Fig.~\ref{Fig:edge}. For example, base stations equipped with a small data center or a set of servers in radio access networks can accept offloaded jobs from adjacent mobile and IoT devices~\cite{YuanWu.Offloading.2017}. By providing local computing power, edge computing enables blockchain deployment in IoT networks to support solving PoW puzzles, hashing, encryption algorithms, and possibly consensus.
		
	Interactions among edge computing service providers and IoT devices or users can be modeled as market activities, where the providers sell resources such as data and computing power, generating a revenue from IoT users. In practice, the similar concept that integrates cloud and blockchain has been realized. For example, Microsoft provides Blockchain as a Service (BaaS) on the Azure cloud platform. A UK company CloudHashing offers Bitcoin Mining as a Service (MaaS) where the users only buy software services online to mine Bitcoins, without installing and deploying hardware equipment. IBM provides Watson IoT platform to manage IoT data in a private blockchain ledger, which is integrated in IBM's business-level cloud services. Despite all the cloud-based blockchain services, economic models for blockchain transactions in edge computing systems are not well studied. Furthermore, the edge computing service is able to provide the near-to-end computing units such as fog nodes or edge devices which are closer to the miners, where the latency is much smaller than that of the cloud computing. Therefore, it is suitable for delay-sensitive IoT applications.
	
	\subsection{Application Scenarios}
	Edge computing services for mobile blockchain can be applied to various application scenarios.
	
		\subsubsection{Secured Smart Home}
		Home appliances that upload a large quantity of home and personal data to centralized databases controlled by smart device manufacturers or retailers may be exposed to serious privacy problems~\cite{Christidis.B4IoT.2016}. Therefore, edge computing for IoT with blockchain can provide a transparent and secured alternative framework for private data management in smart home.
		
		\subsubsection{Smart Grid}
		Smart grid systems~\cite{YuanWu.Smartgrid.2017} contain heterogeneous sensors, e.g., smart meters. For smart meters, energy consumption data and energy transactions are recorded. The edge computing can be integrated alongside with smart meters to process complex jobs automatically, e.g., preparing transactions, executing smart electricity contracts and balancing the grid load~\cite{IBM.2016}. The similar concept can be applied for plug-in hybrid electric vehicles (PHEVs)~\cite{Kang.PHEVBlockchain.2017} to support energy storage sharing.
	
	\subsection{Open Issues}
	Although blockchain has been applied to many applications, some open issues exist.

		\subsubsection{Security}
		Although blockchain offers attractive security features for distributed data processing and storage, some security issues emerge when blockchain is used with edge computing, especially private blockchains. One possible implementation can be the small-scaled ``whitelisted'' IoT networks~\cite{Christidis.B4IoT.2016} where all network nodes trust each other, and PoW is not required. In this case, attacks can happen when transaction data is transferred from mobile or IoT devices to edge computing servers. A secure and trusted network between the devices and servers is required. Distributed Denial of Service (DDoS) attacks can be easily launched to disrupt the connection and operation of mobile blockchain. Likewise, since the mobile and IoT devices rely on wireless transmission, jamming attacks can disrupt the blockchain data exchange. Finally, with PoW, mobile blockchain is subject to typical 51\% attacks that attackers control majority of computing power. This attack becomes likely as the mobile and IoT devices have limited computing power and/or may not rely on external costly computing services.

		\subsubsection{Resource Utilization and Allocation}
		As aforementioned, mobile or IoT users can use edge computing services for the mobile blockchain. However, edge computing services have limited capacity which may not be sufficient to support demand from all users. Edge computing resource allocation issue arises. On the one hand, the users have different valuation of the edge computing services. The valuation depends on certain factors such as reward and number of transactions included in a block of the mobile blockchain. These factors influence the computing resource demand of the users. The demand should be derived to maximize the utility of the users. On the other hand, the users can compete for the edge computing services offered by the provider. Therefore, the provider can maximize its profit by adjusting the price. For example, when the edge computing resource demand is high, the price can be increased. On the contrary, when the demand is low, the price can be decreased. With this fact, an economic model become a suitable approach for the resource allocation.

\section{Prototype of Edge Computing for Mobile Blockchain}\label{Sec:Demo}
	\begin{figure}
	\begin{center} \subfloat[]{\includegraphics[width=.42\textwidth]{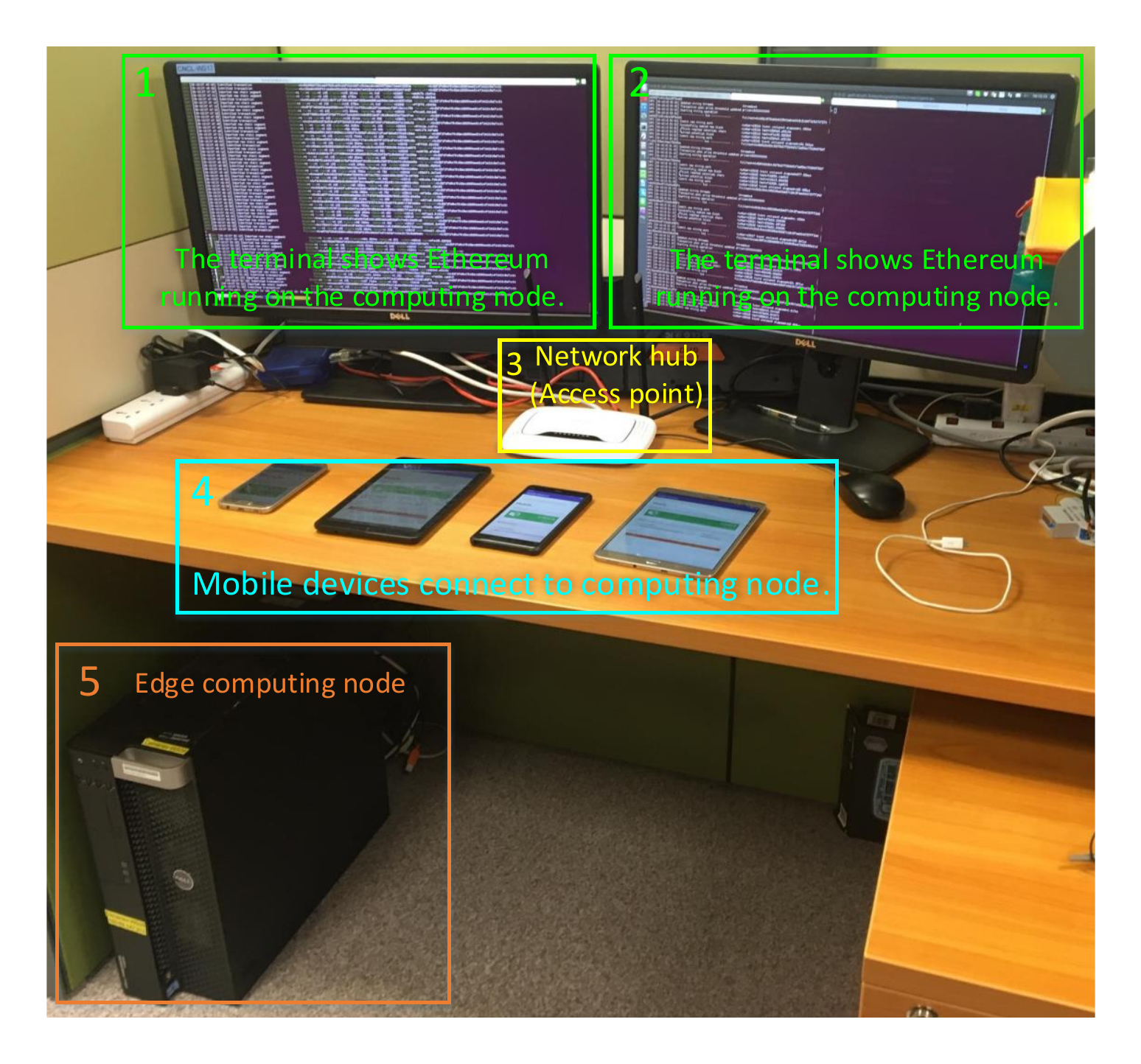}}\hspace{3mm}
	\subfloat[]{\includegraphics[width=.46\textwidth]{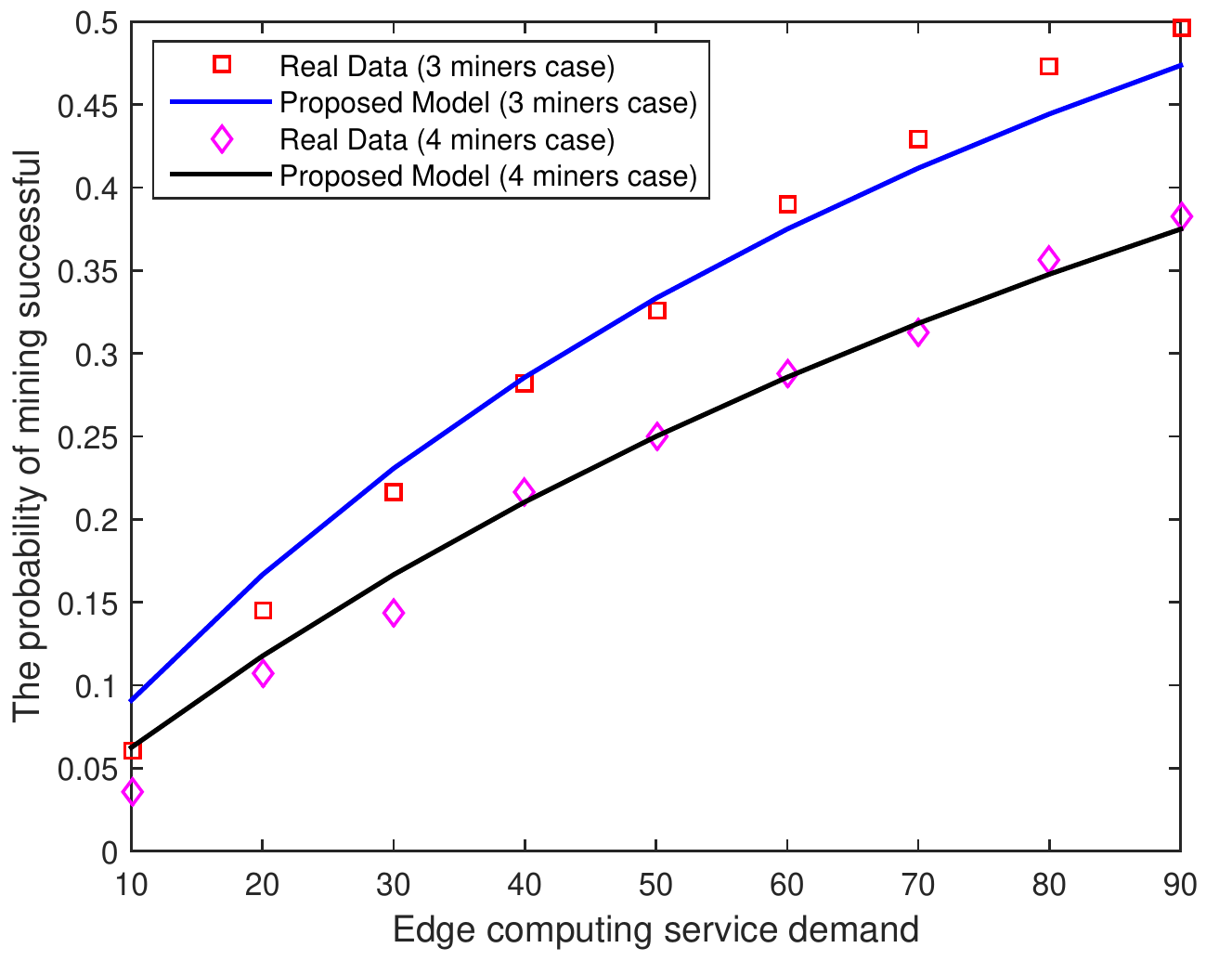}} \\
	\end{center}
	\caption{\small (a) Real mobile blockchain mining experimental setup with Ethereum which is a popular open ledger (b) The comparison of experimental results with analytical results.}
	\label{Fig:Experiment}
	\end{figure}
	
To demonstrate the feasibility and practicability of the edge computing for mobile blockchain presented in Section~\ref{Sec:MECBlockchain}, we implement a prototype system. As illustrated in Fig.~\ref{Fig:Experiment}(a), in the mobile blockchain network, nodes, e.g., mobile or IoT devices, need to perform mining on the edge computing server. The prototype uses a workstation with Intel Xeon CPU E5-1630 as the edge computing server and Android devices as the mobile nodes. The nodes act as the miners, installing a mobile blockchain client application. The application can record data using internal sensors, e.g. an accelerometer and GPS, or the transactions of mobile peer-to-peer data transfer. In Fig.~\ref{Fig:Experiment}(a), the computer screens shown in Boxes $1$ and $2$ display the Ethereum, which is one of the most commonly used blockchain platform, running on the workstation, i.e., Box $5$. The mobile devices in Box $4$ are connected to the edge computing server through a network hub (Box $3$) using the mobile blockchain client application. The basic mining steps can be implemented as follows. The miners request the computing service from the server and mine a block with the assistance of Ethereum services provided accordingly. The mined blocks in blockchain can be accessed and distributed through the Ethereum function.

We conduct the following experiments for performance evaluation. We first create $1000$ blocks using ${\mathrm{Node.js}}$ and use the mobile devices to initiate the mining of the blocks. We consider two cases with three miners and four miners. In the three miner case, we first fix edge computing service demand (CPU utilization) of two miners at $40$ and $60$, and vary the demand of the other miner. In the four miner case, we fix the demand of three miners at $40$, $50$ and $60$, and the demand of the other miner. The number of transactions in each mined block is $10$, i.e., the size of a block is the same. Note that we can control the time-scale of the experiment by adjusting the ``difficulty target'' in solving the PoW, as described in Section~\ref{SubSec:Cons-and-Minn}.

The experimental results are shown in Fig.~\ref{Fig:Experiment}(b) in which when the edge computing service demand increases, the probability that a miner successfully mines a block is higher. We also compare the experimental results with the analytical results as derived in~\cite{Xiong.JSAC.arXiv.2017}. Clearly, the analytical and experimental results match well.

\section{Edge Computing Resource Management for Mobile Blockchain}	\label{Sec:Model}	
In this section, we first present the system model of edge computing for mobile blockchain. Again, the provider deploying the mobile edge computing service aims to maximize the profit through pricing. Therefore, the miners have to consider the reward from mining and the price paid to the provider in deciding on the service demand. Therefore, we present the Stackelberg game formulation and obtain the equilibrium.

	\subsection{Mobile Edge Computing for Blockchain Mining}\label{Subsec:ChainMining}
		\begin{figure}
		\begin{center}
		\subfloat{\includegraphics[width=.40\textwidth]{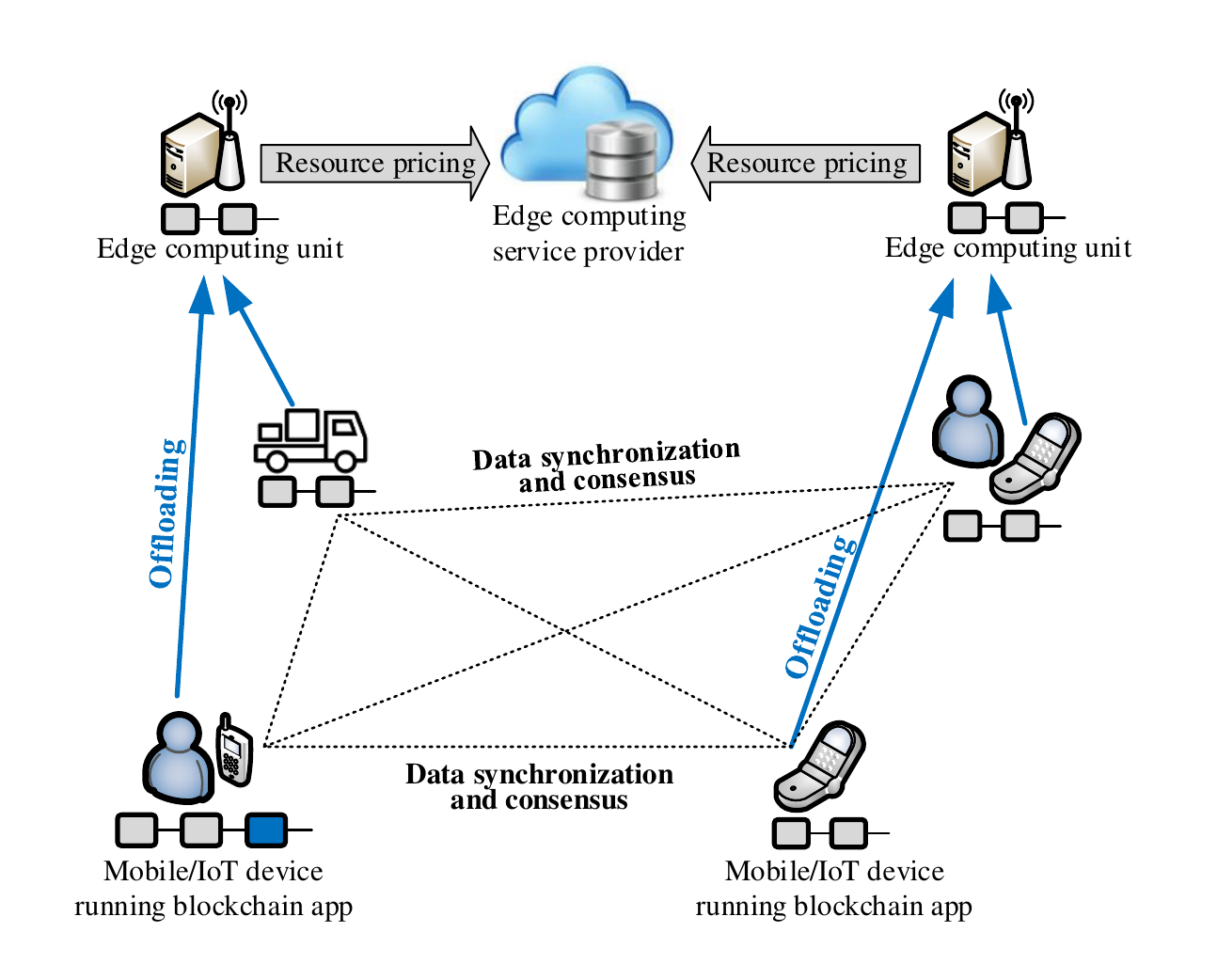}}\hspace{3mm}
		\subfloat{\includegraphics[width=.45\textwidth]{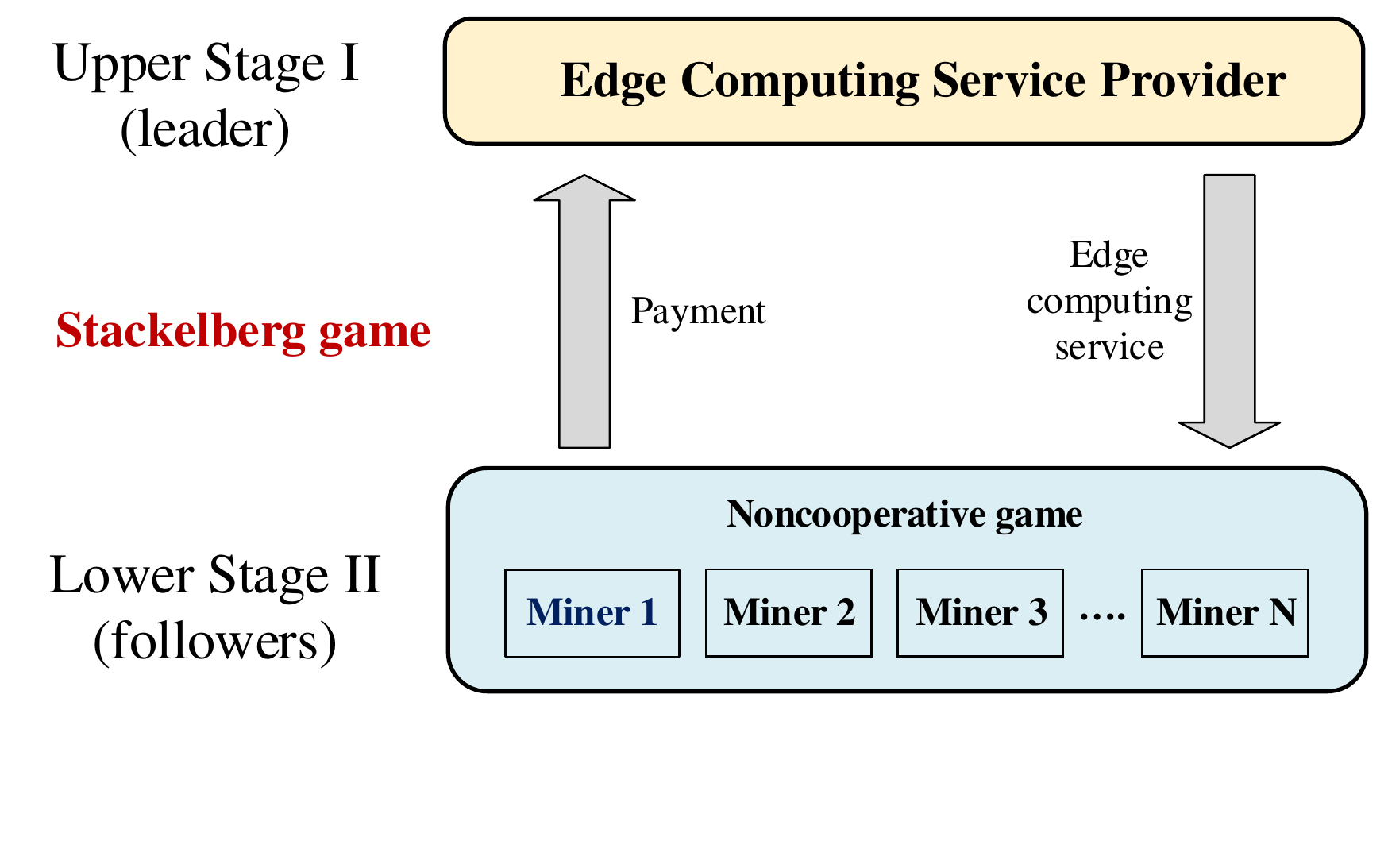}} \\
		\end{center}
		\caption{\small Mobile edge computing enabled blockchain system with the two-stage Stackelberg game model.}
		\label{Fig:Model}
		\end{figure}
	
	The mobile blockchain network is composed of $N$ users acting as the miners. Each user runs mobile blockchain applications to record transaction data. An edge computing service provider deploys edge computing servers for the miners. The PoW puzzle can be offloaded to the edge computing server, and the miners are priced by the provider. Figure~\ref{Fig:Model} shows the system model of the mobile blockchain network. Note that we consider a relatively small blockchain network, where the communication delay due to wireless access is negligible. The offloaded mining process is secured by some security solutions, such as \textit{data masking} or \textit{obfuscation} to hide the details of puzzle solving of the blockchain.

	The provider, i.e., the seller, receives the payment from the miners, i.e., the buyers, which access and consume the edge computing service. The communication and access to edge computing service also consumes energy of miners, which is, nevertheless, much smaller than that needed to solve the PoW. Therefore, a mobile device the user is capable of supporting the communication. Each miner decides on service demand in accordance with the current blockchain operation, e.g., blockchain data synchronization. Here, the demand can be the CPU speed or the number of CPU cores to be used. The provider accepts the demand and the edge computing server processes the offloaded PoW solving for the miners. We can treat the edge computing service demand as the (bought) computing power of the corresponding miner.

	In the mobile blockchain network, miners compete against each other in order to be the first to successfully solve the PoW. Once successfully solving the PoW, the miner broadcasts its solution to the mobile blockchain network to reach consensus. The first miner to successfully mine a block that reaches the consensus earns the mining reward. The reward is from the blockchain applications, expressed as a utility function. The utility function of the $i^{\mbox{th}}$ miner who has $x_i$ edge computing demand consists of two aspects: (1) The reward gained when a block is mined and the consensus is reached, where the reward can include both a fixed and a variable parts; (2) The cost incurred when solving the PoW, i.e., miner $i$ pays a unit price $p_i$ to the provider for using the edge computing service.
	
	\subsection{Two-Stage Stackelberg Game Formulation and Equilibrium}\label{Subsec:GameModel}
	Game theory can be used to analyze the interaction among the provider and miners. For example, in~\cite{Zhang.Fog.2017}, the authors formulated a hierarchical Stackelberg game to solve the resource management in fog computing networks, where the game theoretic study of the market and pricing strategies are proposed. Similarly, the interaction between the provider and miners can be modeled as a Stackelberg game, as illustrated in Fig.~\ref{Fig:Model}. The provider, i.e., the leader of the game, first sets the price of the service per computing unit in the upper Stage I. The miners, i.e., the followers, decide later on their optimal computing service demand for offloading the mining task in the lower Stage II, being aware of the price set by the provider. By using backward induction, we formulate the optimization problems for the leader and followers as two sub-games, as follows.
		\begin{itemize}
			\item {\bf Sub-game (Miners' Mining Strategies in Stage II)}: When participating in the sub-game, each rational miner $i$ decides on the mining demand $x_i$ maximizing the expected utility, given all the other miners' demands, i.e., strategies, as well as the edge computing service price $p_i$.	
			\item {\bf Sub-game (Provider's pricing strategies in Stage I)}: The profit of the provider is the revenue obtained from charging the miners for edge computing service minus the cost. The service cost is a function of the service demand, e.g., electricity consumed.
		\end{itemize}
			
	The two stages of the sub-games together form the Stackelberg game. The objective of the game is to find the Stackelberg equilibrium. The Stackelberg equilibrium ensures that the service price, chosen by the provider, maximize the profit, given that all the miners achieve their best responses which also maximize their utility. The best responses of the miners are obtained given the service price from the provider, and the best responses constitute the Nash equilibrium in the miners' sub-game.
		
	For the provider's pricing, the uniform and discriminatory pricing schemes are adopted. Miners are charged by the provider with the same service price in the uniform pricing scheme. For the discriminatory pricing scheme, the provider charges each miner with exclusive service price. The uniform pricing is easier to implement as the provider does not need to keep track of information of all miners, and charging the same prices is fair for the miners. However, it may not yield the highest profit compared with discriminatory pricing scheme. The reason is that the provider can customize the service price specifically for each miner based on the the miner's preferences in the utility.

	The solution of the Stackelberg game, as well as the proof of the existence and uniqueness of each sub-game equilibrium can be done by employing backward induction. The detailed proof steps are presented in~\cite{Xiong.JSAC.arXiv.2017}. 

\section{Numerical Results}\label{Sec:Numerical}
We investigate the performance of the proposed edge computing resource management for mobile blockchain. We consider a group of mobile blockchain miners in the network and assume the size of a block mined by a miner follows the normal distribution ${\cal N}(\mu_t, \sigma^2)$. The default system parameter values are set as follows: $\mu_t = 200$, $\sigma^2 = 5$, $R = 10^4$, $r = 20$ and $N=100$.

	\subsection{Impacts of the Number of Miners}\label{Subsubsubsec:number}
		\begin{figure}[htbp]
		\begin{center}
		\includegraphics[width=.45\textwidth]{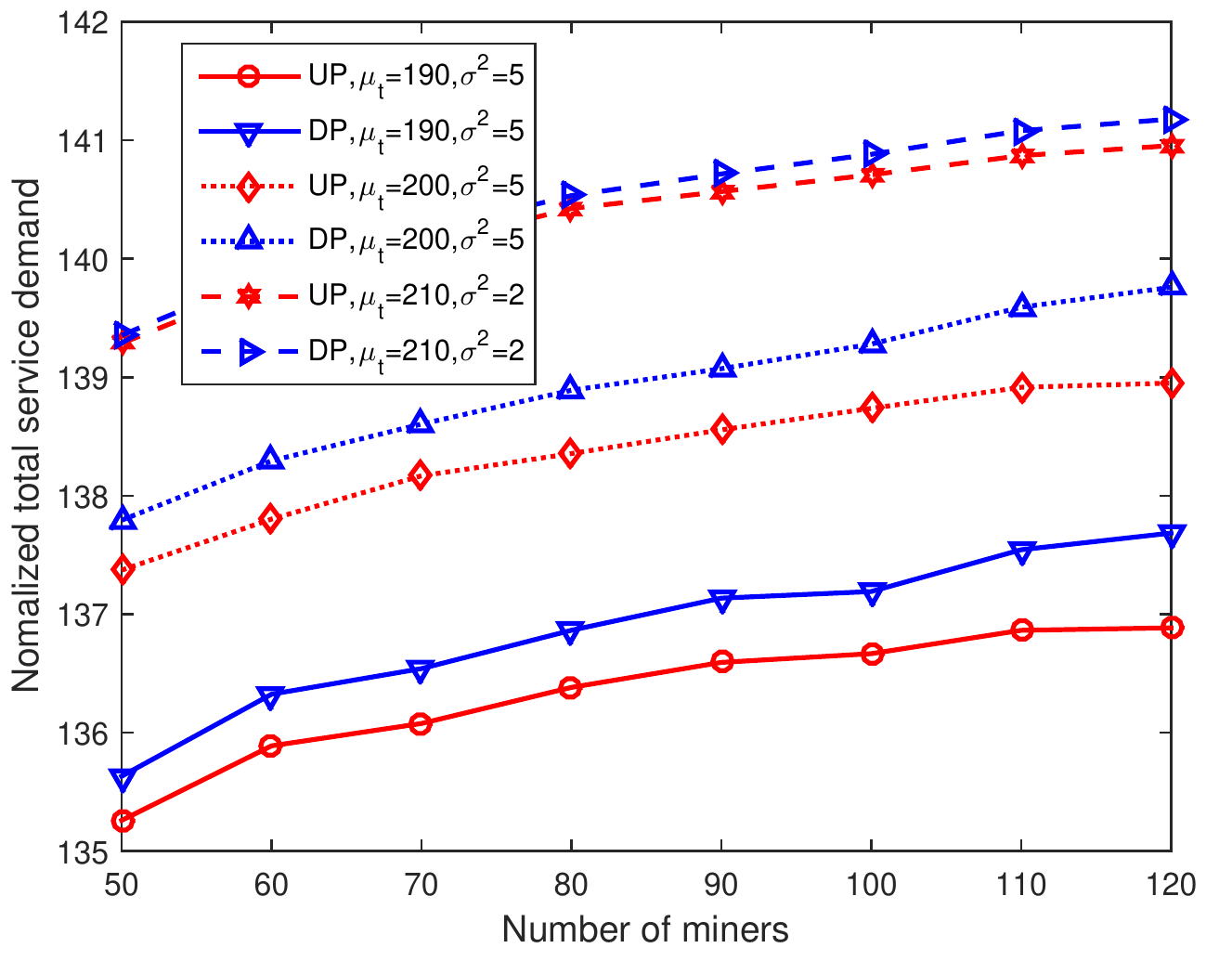}
		\caption{\small Impacts of number of miners on total service demand. (In the legend, UP: Uniform Pricing, and DP: Discriminatory Pricing)}
		\label{Fig:ImpactsUserNumber}
		\end{center}
		\end{figure}
	
	From Fig.~\ref{Fig:ImpactsUserNumber}, the total service demand of miners and the profit of the provider increase with the increased number of miners in mobile blockchain. The reason is that having more miners can generate more demand to the mobile edge computing service. In turn, the provider extracts more surplus from miners and thereby has greater profit gain. Additionally, the service demand increase becomes saturated as the number of miners increases to a certain point. The reason is that the incentive of miners to increase their service demand is weakened because the probability of their successful mining is lower with more miners. Moreover, the total service demand of miners and the profit of the provider increase as $\mu_t$ increases. This is because when $\mu_t$, the average size of one block increases, the variable reward for each miner also increases. The incentive of miners to increase their service demand is improved, and accordingly the total service demand of miners increases. Consequently, the provider achieves greater profit gain. Further, we observe that the total service demand of miners is lower under uniform pricing than that under discriminatory pricing. The intuition is that, under discriminatory pricing, the provider can set different unit prices of service demand for different miners. Slightly lower prices can be charged to encourage higher total service demand from miners. This intuition is consistent with the following results.

	\subsection{Impacts of Reward on Miners}\label{Subsubsec:price}
		\begin{figure}
		\begin{center}
		\subfloat[]{\includegraphics[width=.45\textwidth]{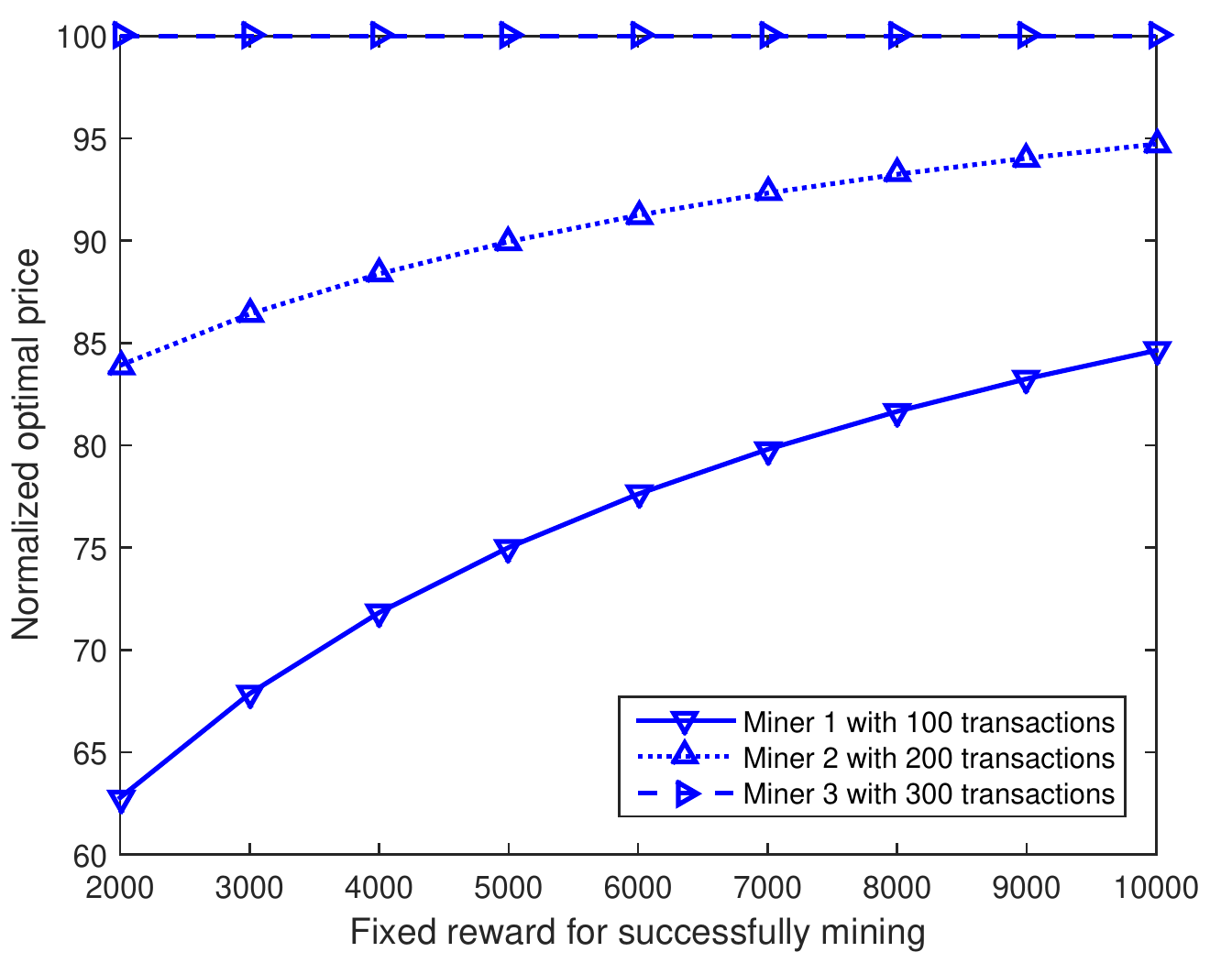}}
		\subfloat[]{\includegraphics[width=.45\textwidth]{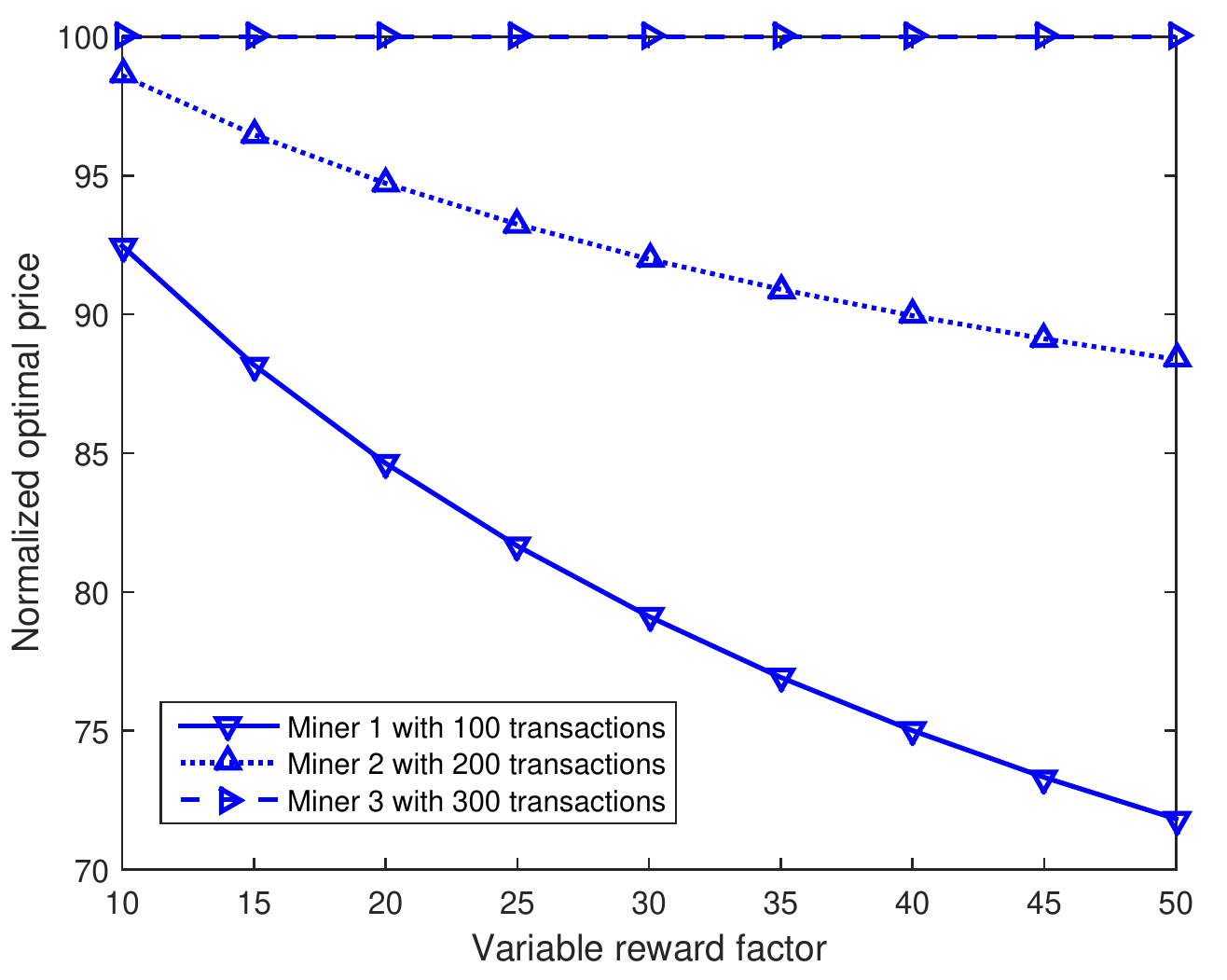}}
		\end{center}
		\caption{\small Impacts of mining reward: (a) Fixed mining reward to optimal edge service price; (b) Variable reward factor to demand of miner. }
		\label{Fig:ImpactsReward}
		\end{figure}
		
	We study a group of $3$ miners in the mobile blockchain network to study the impacts of reward on each specific miner. From Fig.~\ref{Fig:ImpactsReward}(a), the optimal price under discriminatory pricing charging to the miners with the smaller block size is lower, e.g., miners $1$ and $2$. The reason is that the variable reward of miners $1$ and $2$ for successful mining is smaller than that of miner $3$. Thus, miners $1$ and $2$ have no incentive to pay a high price for their service demand as miner $3$ does. In this case, the provider can attract more service demand of miners $1$ and $2$ by setting lower prices. Due to the competition from the other two miners for mining, miner $3$ also has an incentive to increase its service demand. However, due to the high service price, miner $3$ reduces its service demand for saving cost. Nevertheless, the increase in service demand from miners $1$ and $2$ are greater.

	Furthermore, from Fig.~\ref{Fig:ImpactsReward}(a), we observe that the optimal prices for miners $1$ and $2$ increase with the increase of fixed mining reward. This is due to the fact that when the fixed reward increases, the incentives of miners $1$ and $2$ to have higher service demand are greater. The provider thus is able to raise the price and charge more for higher revenue, achieving greater profit. Additionally, we observe from Fig.~\ref{Fig:ImpactsReward}(b) that the optimal prices for miners $1$ and $2$ decrease as the variable reward factor increases. The reason is that when the variable reward increases, an incentive of each miner to have higher service demand is greater. However, the incentives of the miners with smaller block to mine, i.e., miners $1$ and $2$, are still not much as that of miner $3$. The incentives become smaller than that of miner $3$ as the variable reward increases. Therefore, the provider intends to set the lower price for miners $1$ and $2$ which may induce more service demand.

	Note that the Stackelberg game of the edge computing service for mobile blockchain aims at maximizing the profit of the provider. Alternatively, social welfare, i.e., utility of miners, are also important and should be maximized. Auction is a suitable tool to achieve this objective in which some preliminary modeling and results are presented in~\cite{Jiao_auction}

\section{Conclusion}\label{Sec:Conclusion}
In this article, we have introduced edge computing for mobile blockchain applications, especially for IoT blockchain mining tasks offloading with a demonstrated testbed and experimental results. Then, for efficient edge resource management for mobile blockchain, we have presented a Stackelberg game model. We have also conducted the numerical simulations to evaluate the network performance, which may help the edge service providers achieve optimal resource management policy and profit.

\section*{Acknowledgment}
This work was supported in part by National Natural Science Foundation of China (Grant No. 61601336), Hubei Provincial Natural Science Foundation of China (Grant No. 2017CFA012), Singapore MOE Tier 1 under Grant 2017-T1-002-007 RG122/17, MOE Tier 2 under Grant MOE2014-T2-2-015 ARC4/15, NRF2015-NRF-ISF001-2277, EMA Energy Resilience under Grant NRF2017EWT-EP003-041, US NSF CNS-1717454, CNS-1731424, CNS-1702850, CNS-1646607, and ECCS-1547201.


\begin{IEEEbiography}[{\includegraphics[width=1in,height=1.25in,clip,keepaspectratio]{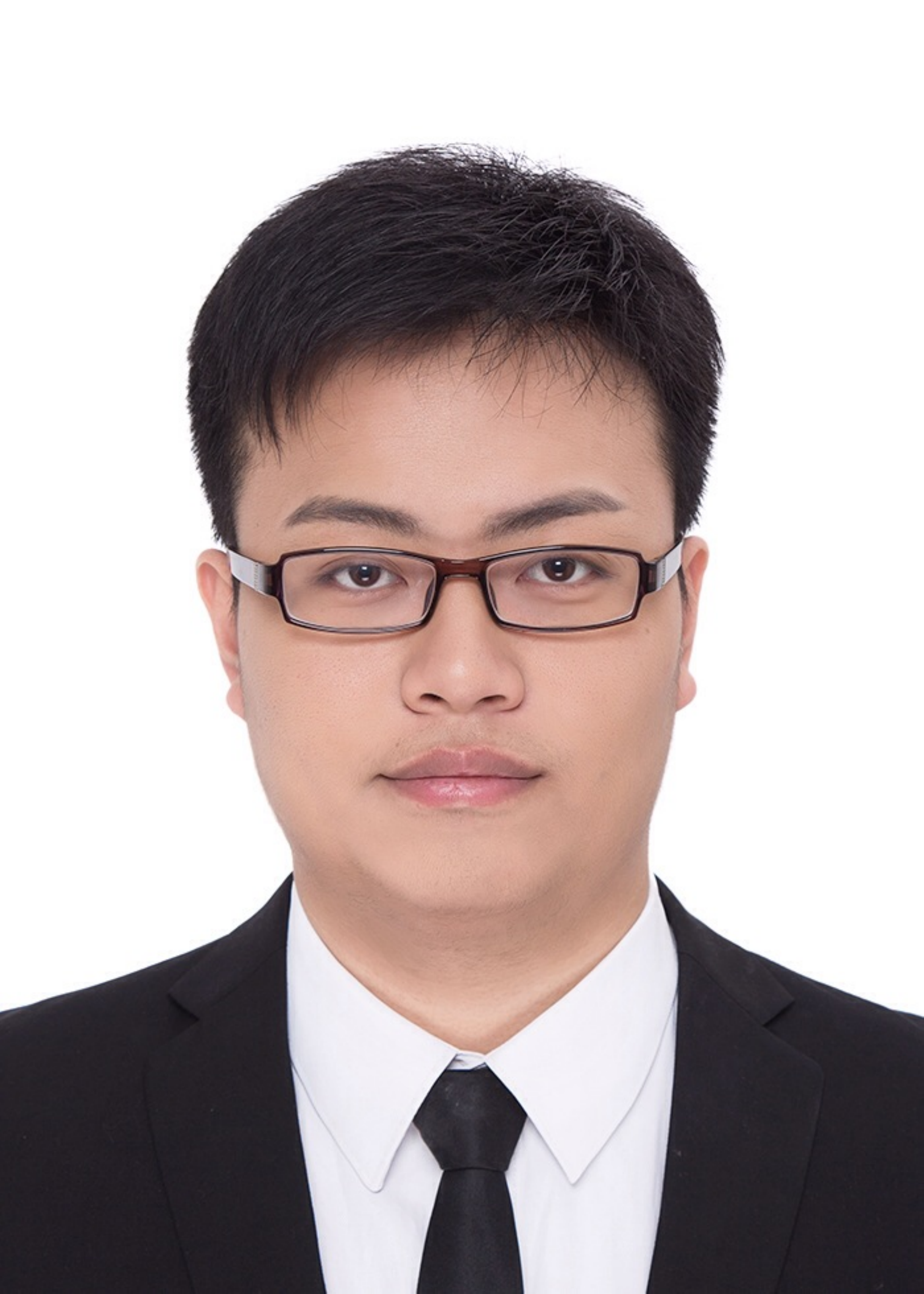}}]{Zehui Xiong}
(S'17) received his B.Eng degree with honors in Telecommunication Engineering from Huazhong University of Science and Technology, Wuhan, China, in 2016. He is currently working towards the Ph.D. degree in the School of Computer Science and Engineering, Nanyang Technological University, Singapore. His research interests include network economics, game theory for resource management, market models and pricing.
\end{IEEEbiography}

\begin{IEEEbiography}[{\includegraphics[width=1in,height=1.25in,clip,keepaspectratio]{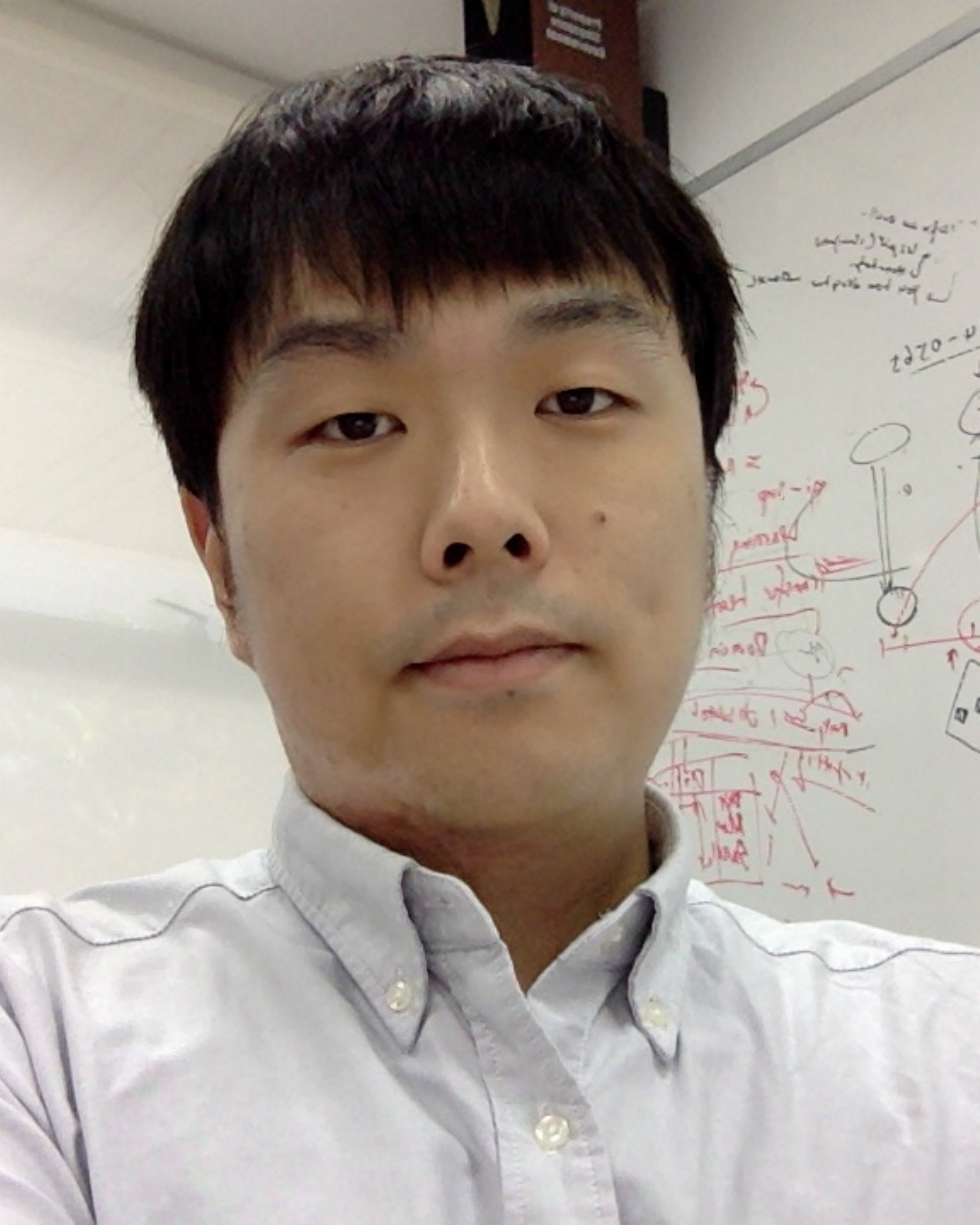}}]{Yang Zhang}
(M'11) is currently an Associate Professor in the School of Computer Science and Technology, Wuhan University of Technology. He received the Ph.D. degree from the School of Computer Science and Engineering, Nanyang Technological University (NTU), Singapore, in 2015. He obtained his B.Eng. and M.Eng. degrees from Beihang University (BUAA), Beijing, China, in 2008 and 2011, respectively. His research interests are market-oriented modeling for wireless resource allocations, and optimal decision making problems.
\end{IEEEbiography}

\begin{IEEEbiography}[{\includegraphics[width=1in,height=1.25in,clip,keepaspectratio]{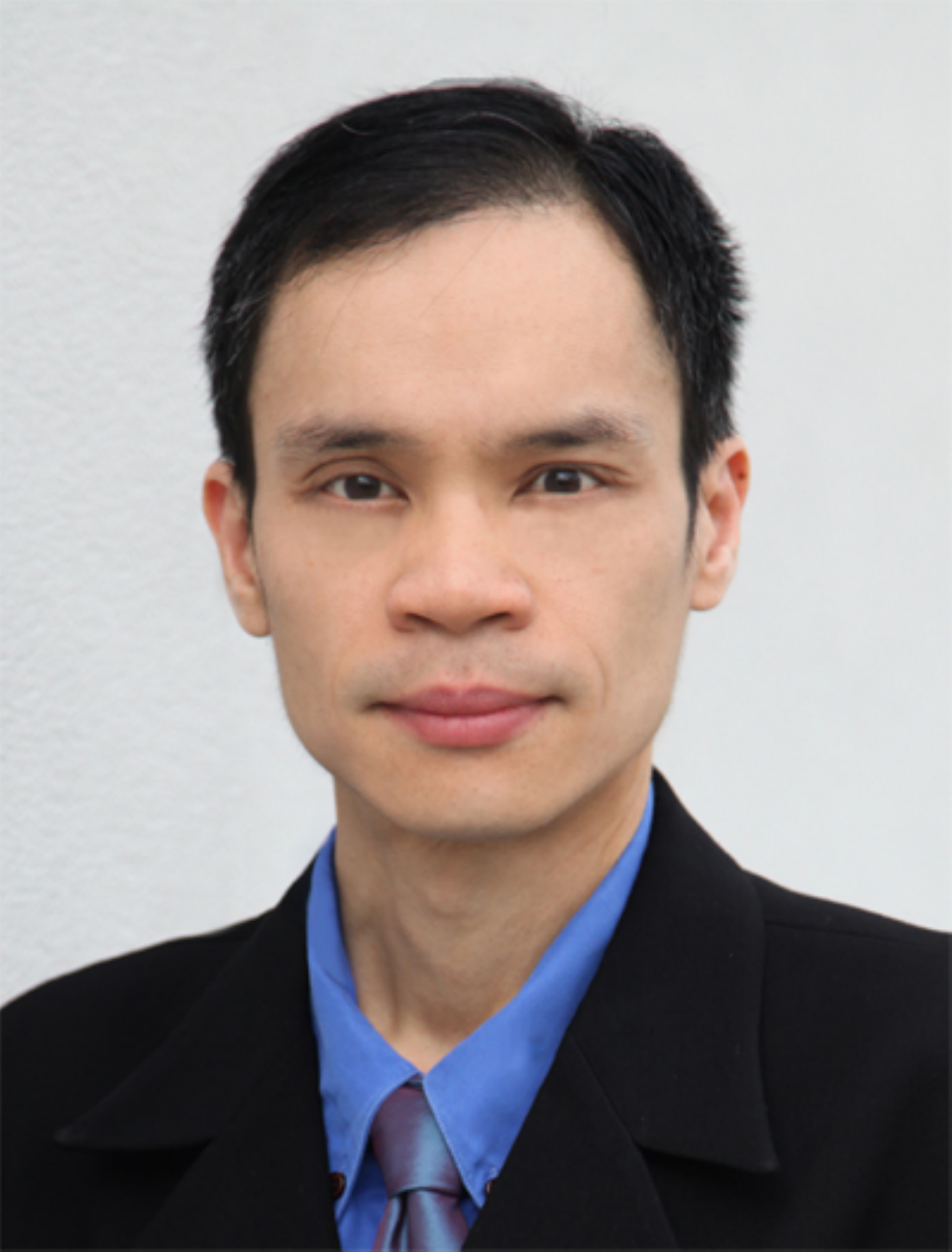}}]{Dusit Niyato}
(M'09-SM'15-F'17) is currently a professor in the School of Computer Science and Engineering, at Nanyang Technological University, Singapore. He received B.Eng. from King Mongkut's Institute of Technology Ladkrabang (KMITL), Thailand in 1999 and Ph.D. in Electrical and Computer Engineering from the University of Manitoba, Canada in 2008. His research interests are in the area of energy harvesting for wireless communication, Internet of Things (IoT) and sensor networks.
\end{IEEEbiography}

\begin{IEEEbiography}[{\includegraphics[width=1in,height=1.25in,clip,keepaspectratio]{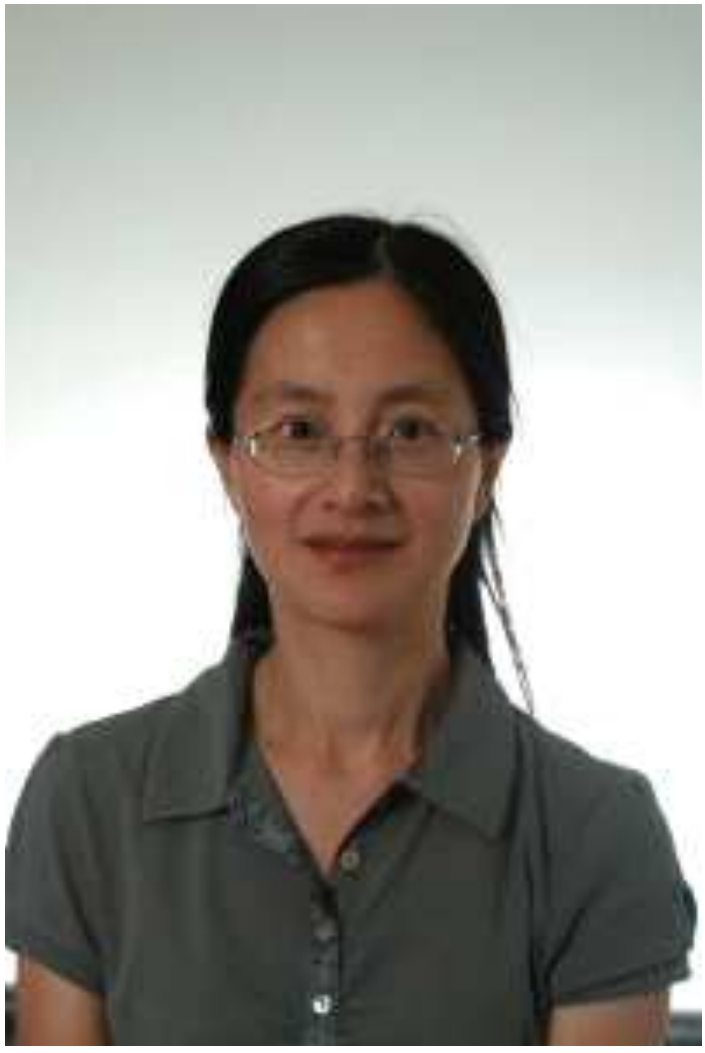}}]{Ping Wang}
(M'08-SM'15) received PhD degree in electrical engineering from University of Waterloo, Canada, in 2008. She is an Associate Professor in the School of Computer Science and Engineering, Nanyang Technological University, Singapore. She was a corecipient of the Best Paper Award from IEEE WCNC 2012 and ICC 2007. She was an Editor of IEEE Transactions on Wireless Communications, EURASIP Journal on Wireless Communications and Networking, and International Journal of Ultra Wideband Communications and Systems.
\end{IEEEbiography}

\begin{IEEEbiography}[{\includegraphics[width=1in,height=1.25in,clip,keepaspectratio]{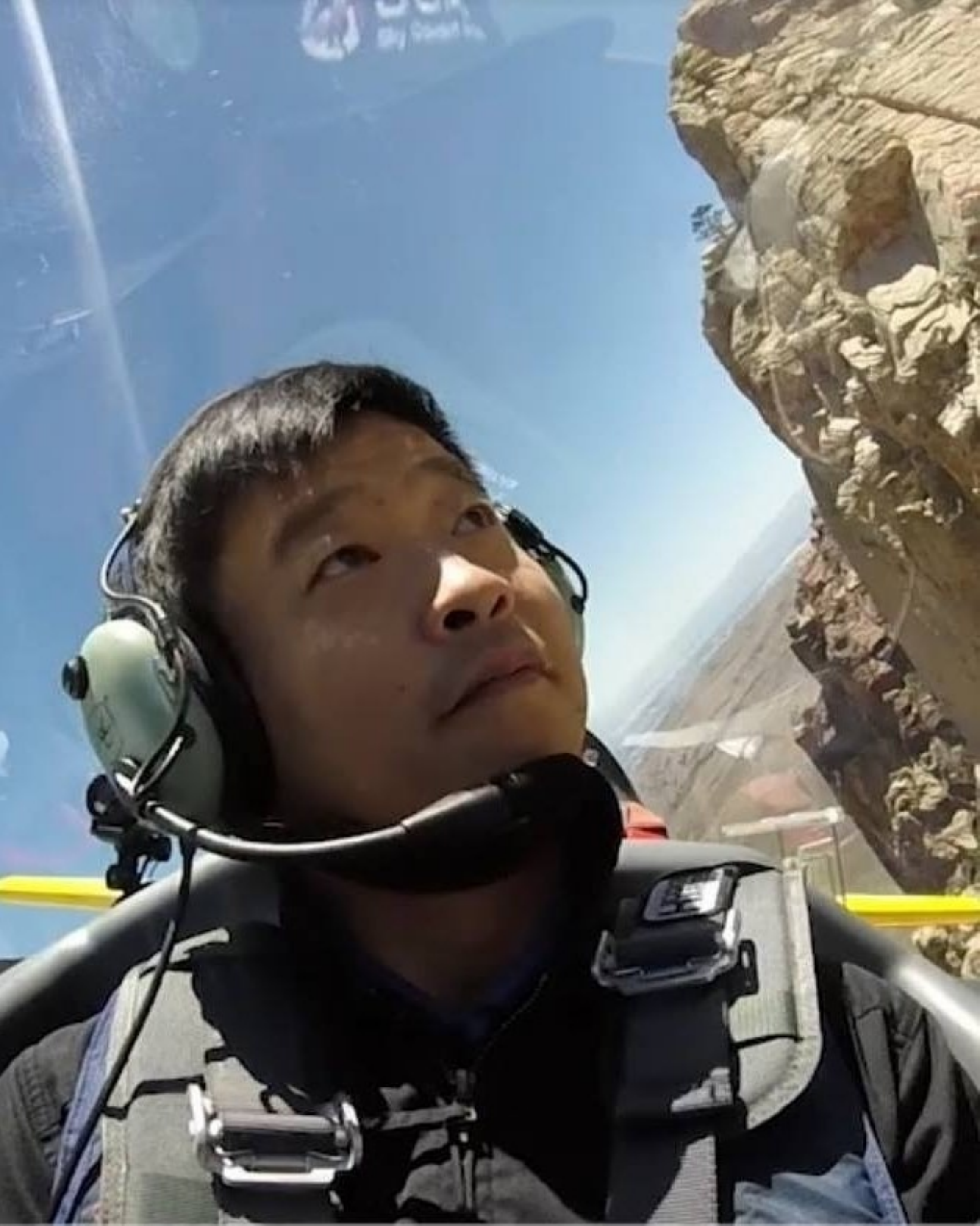}}]{Zhu Han}
(S'01-M'04-SM'09-F'14) received Ph.D. degree in electrical and computer engineering from the University of Maryland, College Park. Dr. Han received an NSF Career Award in 2010, the Fred W. Ellersick Prize of the IEEE ComSoc in 2011, and IEEE Leonard G. Abraham Prize in Communications Systems in 2016. Currently, Dr. Han is an IEEE ComSoc Distinguished Lecturer and 1\% highly cited researcher according to WoS 2017.
\end{IEEEbiography}

\end{document}